\documentclass[conference]{IEEEtran}
\IEEEoverridecommandlockouts
\usepackage{cite}
\usepackage{amsmath,amssymb,amsfonts}
\usepackage{algorithmic}
\usepackage{graphicx}
\usepackage{textcomp}
\usepackage{xcolor}
\usepackage{comment}
\def\BibTeX{{\rm B\kern-.05em{\sc i\kern-.025em b}\kern-.08em
    T\kern-.1667em\lower.7ex\hbox{E}\kern-.125emX}}
\begin{document}

\pdfinfo{
/Title (The Memecoin Phenomenon: An In-Depth Study of Solana's Blockchain Trends)
}
\title{The Memecoin Phenomenon:\\An In-Depth Study of Solana's Blockchain Trends}

 \author{
 \IEEEauthorblockN{Davide Mancino}
 \IEEEauthorblockA{\textit{University of Milano-Bicocca}\\
    \textit{Department of Informatics, Systems, and Communication (DISCo)} \\
    Viale Sarca 336, Edificio U14, 20126 Milano, Italy \\
    \texttt{davide.mancino@unimib.it}}
 }

\maketitle

\begin{abstract}
This paper analyzes the emerging memecoin phenomenon on the Solana blockchain, focusing on the Pump.fun platform during Q4 2024. Using on-chain data, it is explored how retail-focused token creation platforms are reshaping blockchain ecosystems and influencing market participation. This study finds that Pump.fun accounted for up to 71.1\% of all tokens minted on Solana and contributed 40–67.4\% of total DEX transactions. Despite this activity, fewer than 2\% of tokens successfully transitioned to major decentralized exchanges, highlighting a highly speculative market structure. The platform experienced rapid growth, with daily active users rising from 60,000 to peaks of 260,000, underscoring strong retail adoption. This reflects a broader shift towards accessible, socially-driven market participation enabled by memecoins. However, while memecoins lower entry barriers and encourage retail engagement, they introduce significant risks. The volatile and speculative nature of these platforms raises concerns about long-term sustainability and the resilience of the blockchain ecosystem. These findings reveal the dual impact of memecoins: they democratize token creation and alter market dynamics but may jeopardize market efficiency and stability. This paper highlights the need to critically assess the implications of retail-driven speculative trading and its potential to disrupt emerging blockchain economies.
\end{abstract}


\section{Introduction} \label{sec:introduction}

The blockchain ecosystem has witnessed several transformative waves of innovation since its inception. DeFi protocols~\cite{ozili:defi:jbft:2022}, Blockchain Online Social Media (BOSM)~\cite{guidi:bosm:acm:2021, striking_Mancino}, Maximal Extractable Value (MEV) and its implications~\cite{mancino:mev_eth:itasec:2023} and more recently, metaverse technologies have been explored~\cite{gosh:metaverse:ieee:2024}.
Against this backdrop, a new phenomenon has emerged: the rise of retail-focused token creation platforms, particularly in the memecoin sector. These platforms represent a synthesis of previous trends, combining DeFi's accessibility, BOSM's social dynamics, and elements of virtual world engagement. This evolution raises important questions about speculative behavior, and the democratization of blockchain-based financial instruments.

This paper examines this phenomenon through an empirical analysis of \textit{Pump.fun}, a leading memecoin platform on the \textit{Solana} blockchain. By analyzing platform metrics, user behavior, and market dynamics during Q4 2024, the aim is to understand how these new platforms are reshaping blockchain ecosystems and influencing user participation in cryptocurrency markets. This research contributes to the growing body of literature on blockchain market dynamics by examining how retail-focused platforms are creating new paradigms for token creation, trading, and value extraction.

The remainder of this paper is organized as follows: Section \ref{sec:solana} provides background information on the Solana blockchain, while Section \ref{sec:memecoins} explores the concept of memecoins and their significance. Section \ref{sec:pumpfun} introduces Pump.fun and its role in the ecosystem. Section \ref{sec:analysis} presents the detailed analysis of market behavior and dynamics. Section \ref{sec:speculation} examines economic implications and speculative behavior. Finally, Section \ref{sec:related_work} describes related works and Section \ref{sec:conclusion} concludes the article with key insights and future research directions.

\section{Solana Blockchain} \label{sec:solana}
Solana's blockchain architecture is distinguished by its high throughput and low latency, achieved through a unique consensus mechanism that combines \textit{Proof-of-History} (PoH) with \textit{Proof-of-Stake} (PoS). PoH serves as a cryptographic clock, allowing nodes to agree on the time order of events without extensive communication, thereby enhancing efficiency and scalability. This innovation allows Solana to process thousands of transactions per second, significantly outperforming many other blockchains~\cite{yakovenko:solana_whitepaper:onlinedocs:2017}.

Positioned as a high-performance blockchain, Solana addresses scalability issues that have hindered other blockchains. Its architecture supports decentralized applications (dApps) and smart contracts, attracting a diverse range of projects and users. The low transaction fees and rapid processing times of the platform have solidified its reputation as a viable solution for various blockchain applications~\cite{mishra:solana_review:ijict:2024}.

The \textit{Solana Program Library} (SPL) Token program is Solana's on-chain standard for creating and managing tokens, both fungible and non-fungible~\cite{solana:spl_docs:onlinedocs:2023}. It provides a comprehensive framework that ensures compatibility across the Solana ecosystem, enabling high-performance and secure token operations that leverage Solana's speed and parallel processing capabilities. A key feature of SPL tokens is their support for atomic transfers, which allows complex token interactions within a single transaction.

\section{Memecoins} \label{sec:memecoins}
Memecoins are cryptocurrencies inspired by internet memes or trends, often characterized by their humorous nature and the enthusiastic communities that support them. Unlike traditional cryptocurrencies, which may offer specific utilities or technological advancements, memecoins typically lack inherent value or utility, relying instead on their viral appeal and community engagement. 

A notable example is Dogecoin (DOGE), created in December 2013 by software engineers Billy Markus and Jackson Palmer as a joke, inspired by the viral ``Doge'' meme featuring a Shiba Inu dog~\cite{chohan2021dogecoin}. Despite its origins, Dogecoin has gained a dedicated following and has been adopted as a payment method by some companies~\cite{cointelegraph2023dogecoin}. 

Similarly, Shiba Inu (SHIB) emerged as a memecoin aiming to capitalize on the popularity of Dogecoin. It has developed a loyal community and has seen substantial market activity, with its value largely driven by social media hype and community engagement~\cite{cointelegraph2023dogecoinvsshibainu}.

In 2023, Pepe Coin (PEPE), inspired by the meme ``Pepe the Frog'', gained attention in the cryptocurrency market. Embracing its identity as a memecoin, PEPE reached a market cap of \$1.6 billion in mid-2023, sparking a memecoin frenzy~\cite{crypto_news_pepe_2025}.

The memecoin phenomenon has been analyzed in various studies, highlighting the cultural and economic impacts of memecoins within the blockchain ecosystem. For instance, Krause (2024)~\cite{krause:memecoin:ssrn:2024} examines the socio-economic implications of memecoins, emphasizing their role in digital culture and speculative investment behaviors. Long (2024)~\cite{long:multimodal_memecoin:arxiv:2024} explores the multimodal nature of memecoins, analyzing how visual and textual elements contribute to their virality and market performance. Additionally, Kim (2023)~\cite{kim:twitter_memecoin:itp:2023} investigates the influence of social media platforms, particularly Twitter, on the proliferation and market dynamics of memecoins, demonstrating the significant impact of online communities in shaping the memecoin landscape. 

These studies collectively provide a comprehensive understanding of the memecoin phenomenon, shedding light on the interplay between internet culture, social media, and cryptocurrency markets.

\section{Pump.fun: A Memecoin Platform on Solana} \label{sec:pumpfun}
Pump.fun is a platform built on the Solana blockchain launched in January 2024 that enables users to create and trade tokens instantly. This user-friendly interface has democratized token creation, allowing anyone to launch a memecoin within minutes~\cite{decrypt:pumpfun:onlinedoc:2024}.

A distinctive feature of Pump is its ``graduated tokens'' system. Tokens created on the platform can ``graduate'' to decentralized exchanges like Raydium, where they become available for broader trading. This graduation process involves meeting specific criteria, such as achieving a certain level of trading volume and community engagement, which helps filter out low-quality tokens and promotes more promising projects~\cite{webopedia:pumpfun:onlinedocs:2024}. 

Pump also ensures token safety by preventing rug pulls through a fair-launch mechanism. All tokens created on the platform undergo a launch process with no presale or team allocation. Users can buy tokens directly on the bonding curve (a pricing model where the token price increases as more tokens are purchased) and sell at any time to realize profits or losses. Once the market cap reaches \$100k, \$17k of liquidity is deposited into Raydium and permanently burned, promoting scarcity and price stability~\cite{pump_fun_board}.

Since its inception, Pump.fun has experienced significant growth. By November 2024, the platform had facilitated the creation of over 3 million tokens, averaging seven new tokens per minute, reflecting the platform's popularity and the hype around the memecoin trend~\cite{solanafloor:pumpfun:onlinedocs:2024}.

Pump.fun has been a significant driver of the memecoin trend in 2024. Its ease of use and accessibility have lowered the barriers to entry for token creation, leading to an explosion of memecoins and contributing to the broader popularity of these digital assets within the cryptocurrency market~\cite{theblock:memecoin:onlinedocs:2024}.

However, the platform has not been without controversy. In late 2024, Pump.fun faced criticism for hosting livestreams where token creators engaged in extreme actions to promote their tokens. This led to the indefinite suspension of the livestream feature following community backlash~\cite{pump_fun_livestream_backlash}.

Despite these challenges, Pump.fun remains a pivotal platform in the memecoin ecosystem, exemplifying the dynamic and rapidly evolving nature of blockchain-based digital assets.

\section{Analysis of Memecoin Behavior and Market Dynamics} \label{sec:analysis}

This section focuses on detailing the data utilized and the analyses performed.

\subsection{Data Collection and Methodology} \label{subsec:data}
This research presents an analysis of Pump.fun's market dynamics during the fourth quarter (Q4) of 2024, spanning October through December. The selection of this time period is particularly significant as it captures both the platform's explosive growth phase and its subsequent maturation, providing valuable insights into the evolution of retail-focused token creation platforms.

\textit{Dune} \cite{dune} was chosen as the primary data source due to its comprehensive coverage of the Solana blockchain. This analysis focuses on three data tables, each offering unique perspectives on the ecosystem's dynamics.

The first is the \texttt{tokens\_solana.transfers} table, which tracks all token transfer events on the Solana blockchain. Provide crucial information, including sender and receiver addresses, transaction amounts, and timestamps. This data allows us to follow the entire lifecycle of tokens, from their creation through various trading phases, and to their listing on major decentralized exchanges (DEXes).

Next, the \texttt{dex\_solana.trades} table, offers comprehensive details about trading activities across all decentralized exchanges on Solana. It captures key trade-level data such as exact amounts, prices, and the specific DEX programs involved.

The third essential component is the \texttt{solana.instruction\_calls} table. This table records all program interactions on the Solana blockchain.

\subsection{Platform Dominance in Token Creation}
\begin{figure}[!ht]
    \centering
    \includegraphics[width=\linewidth]{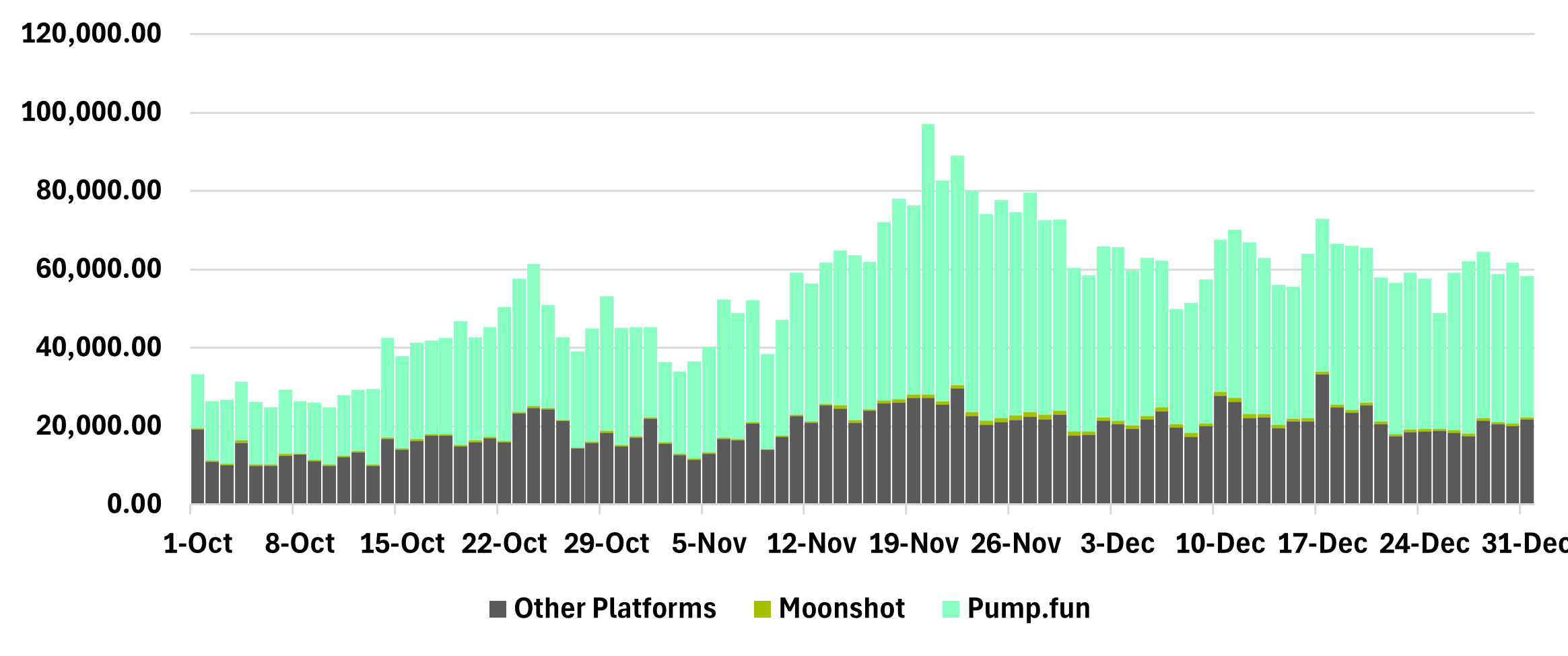} % Sostituisci con il tuo file immagine
    \caption{Daily token creation by various platforms on Solana.} \label{fig:token}
    
    \vspace{0.5cm} % Spazio tra le immagini
    
    \includegraphics[width=\linewidth]{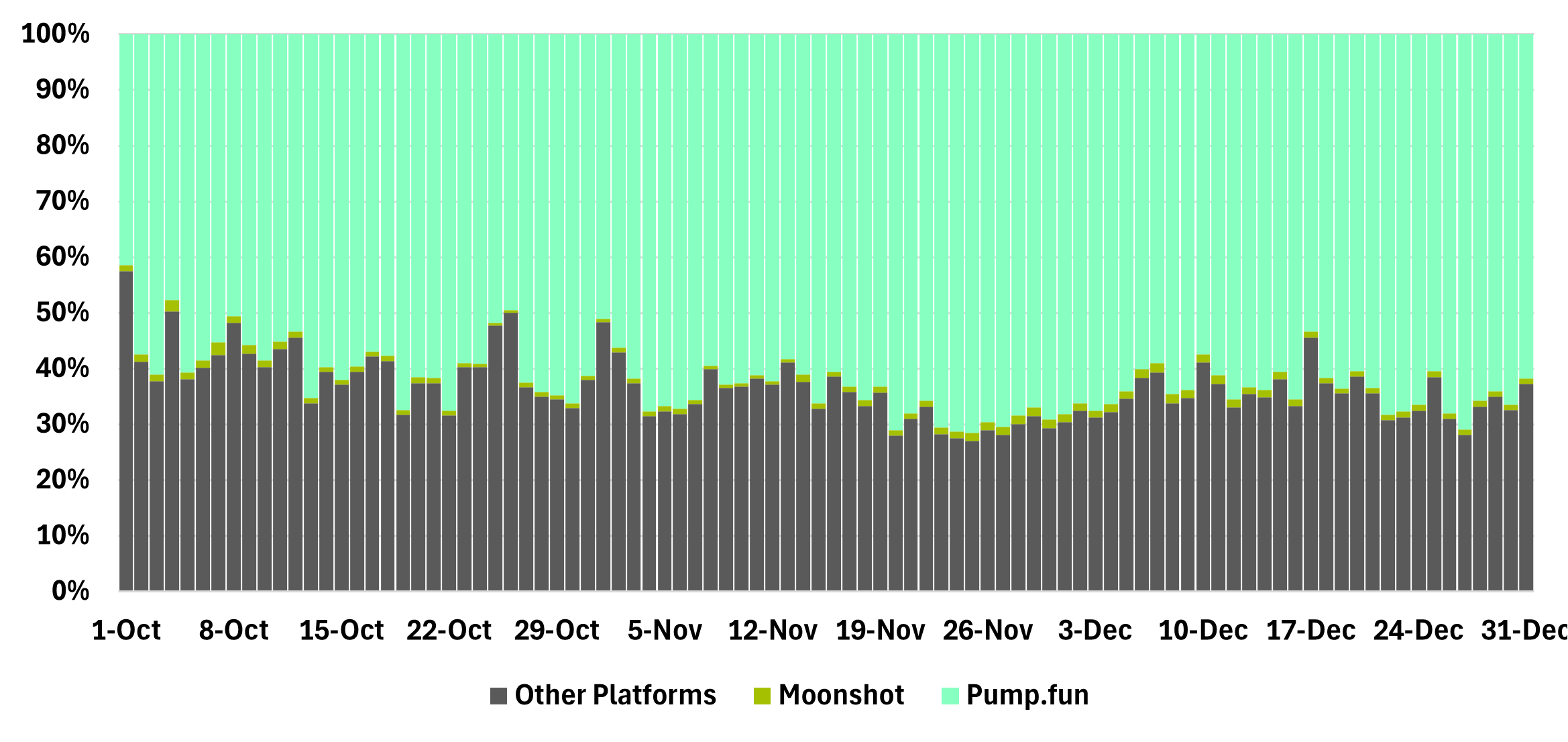} % Sostituisci con il tuo file immagine
    \caption{Percentage of daily token creation by various platforms on Solana.} \label{fig:percToken}
\end{figure}

This analysis aims to understand platform dominance in the creation of new tokens.  
The data used in this study was extracted using SQL queries executed in the Dune Analytics Solana blockchain token transfer database. Minting transactions related to Pump.fun and Moonshot (another platform for buying and selling memecoins) \cite{moonshot} were identified by filtering for their respective executing accounts (\textit{outer\_executing\_account}). The query aggregated the daily count of minted tokens between October 1, 2024, and December 31, 2024. Tokens minted by other platforms were gathered by excluding minting transactions associated with Pump.fun and Moonshot.  
The results were combined to provide a comparative overview, underscoring Pump.fun's dominance over competing platforms.  
A thorough examination of the data, visualized in Figure~\ref{fig:token} and Figure~\ref{fig:percToken}, reveals that Pump.fun has achieved unprecedented superiority in token creation compared to its competitors, including other memecoin platforms such as \textit{Moonshot}.  

The platform's dominance is most strikingly illustrated by its peak performance of 69,046 token mints in a single day, representing an extraordinary 71.1\% of all tokens minted on the Solana blockchain during that period.

\subsection{DEX Transaction Analysis}
\begin{figure}[!ht]
    \centering
    \includegraphics[width=\linewidth]{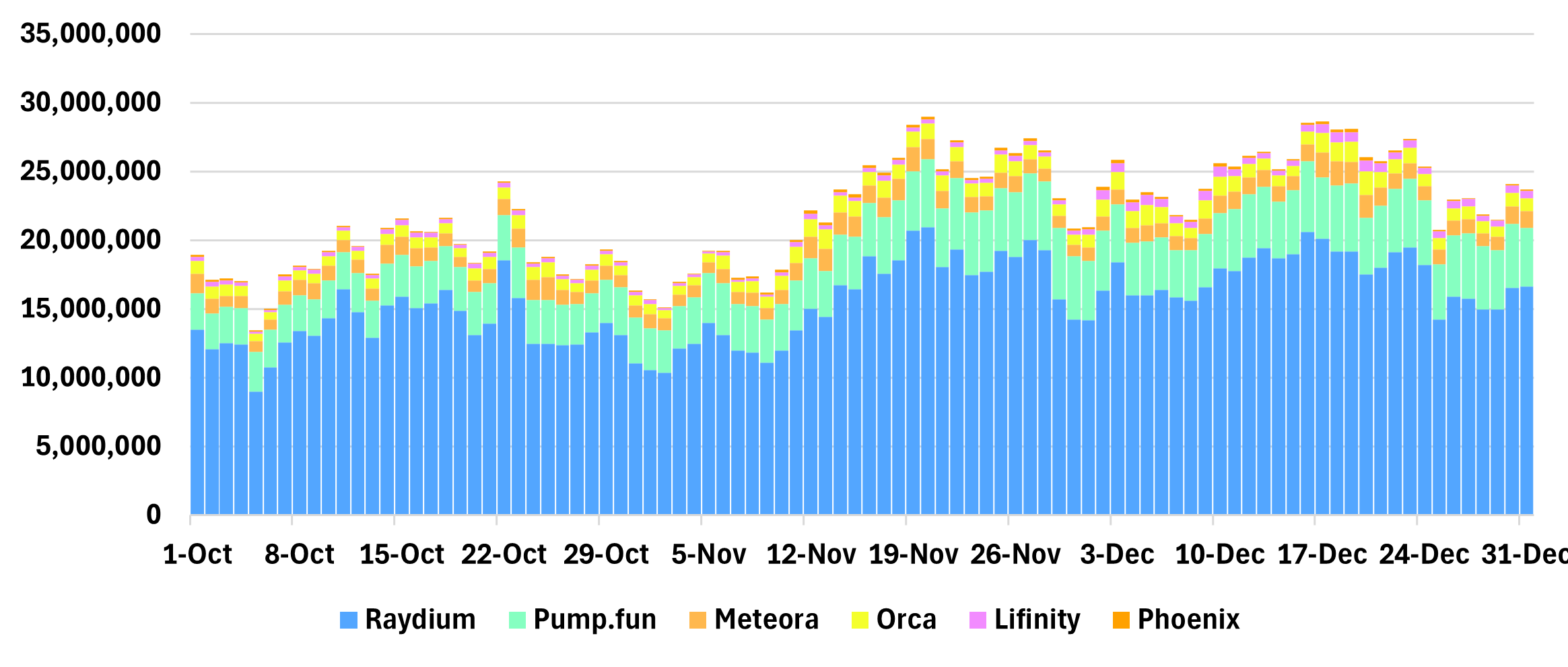} % Sostituisci con il tuo file immagine
    \caption{DEX daily transactions.} \label{fig:dex}
    
    \vspace{0.5cm} % Spazio tra le immagini
    
    \includegraphics[width=\linewidth]{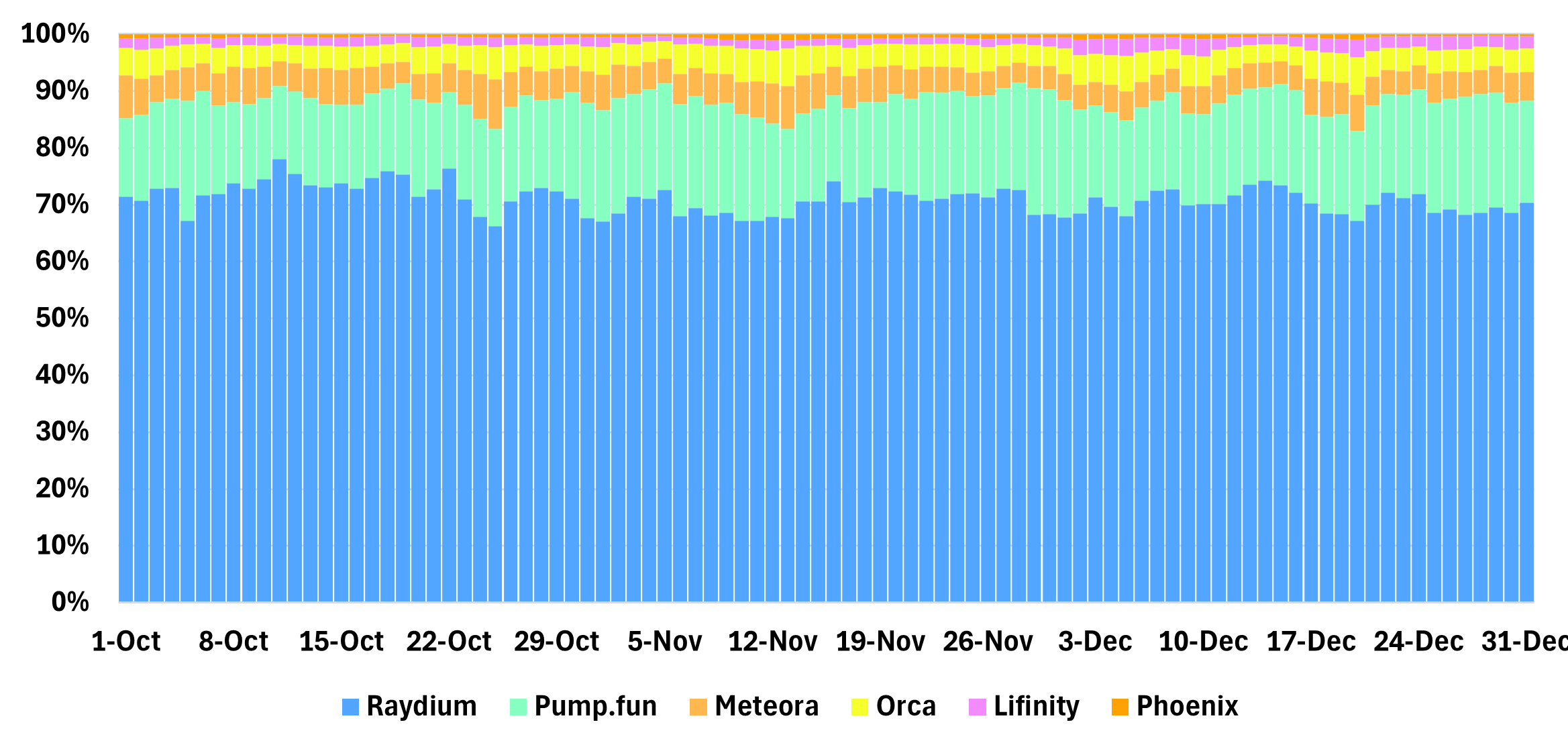} % Sostituisci con il tuo file immagine
    \caption{Percentage of DEX daily transactions.} \label{fig:percDEX}
\end{figure}

This section investigates Pump.fun's competitive standing within the Solana ecosystem by analyzing decentralized exchange (DEX) transaction data. Specifically, the Solana DEX platforms analyzed include Raydium \cite{raydium}, Meteora \cite{meteora}, Orca \cite{orca}, Lifinity \cite{lifinity}, Phoenix \cite{phoenix}, and Pump.fun.
The data, visualized in Figure~\ref{fig:dex} and Figure~\ref{fig:percDEX}, highlights Pump.fun’s strong market presence despite its relatively recent entry.

The data for this analysis was extracted using SQL queries executed on the Dune Analytics Solana blockchain DEX trade database. The query identifies DEX trades conducted between October 1, 2024, and Decemebre 31, 2024, aggregating the daily number of transactions by project. 

Pump.fun consistently accounts for 15\% to 25\% of total DEX activity, a notable achievement considering the competition from long-established exchanges. 

Daily transactions serves as a critical indicator of Pump.fun’s market penetration, with the platform consistently processing between 2 and 4 million transactions each day. This level of activity further solidifies its position among the top-tier exchanges within the Solana ecosystem.  

\subsection{Pump.fun Related Transactions}
\begin{figure}[!ht]
    \centering
    \includegraphics[width=\linewidth]{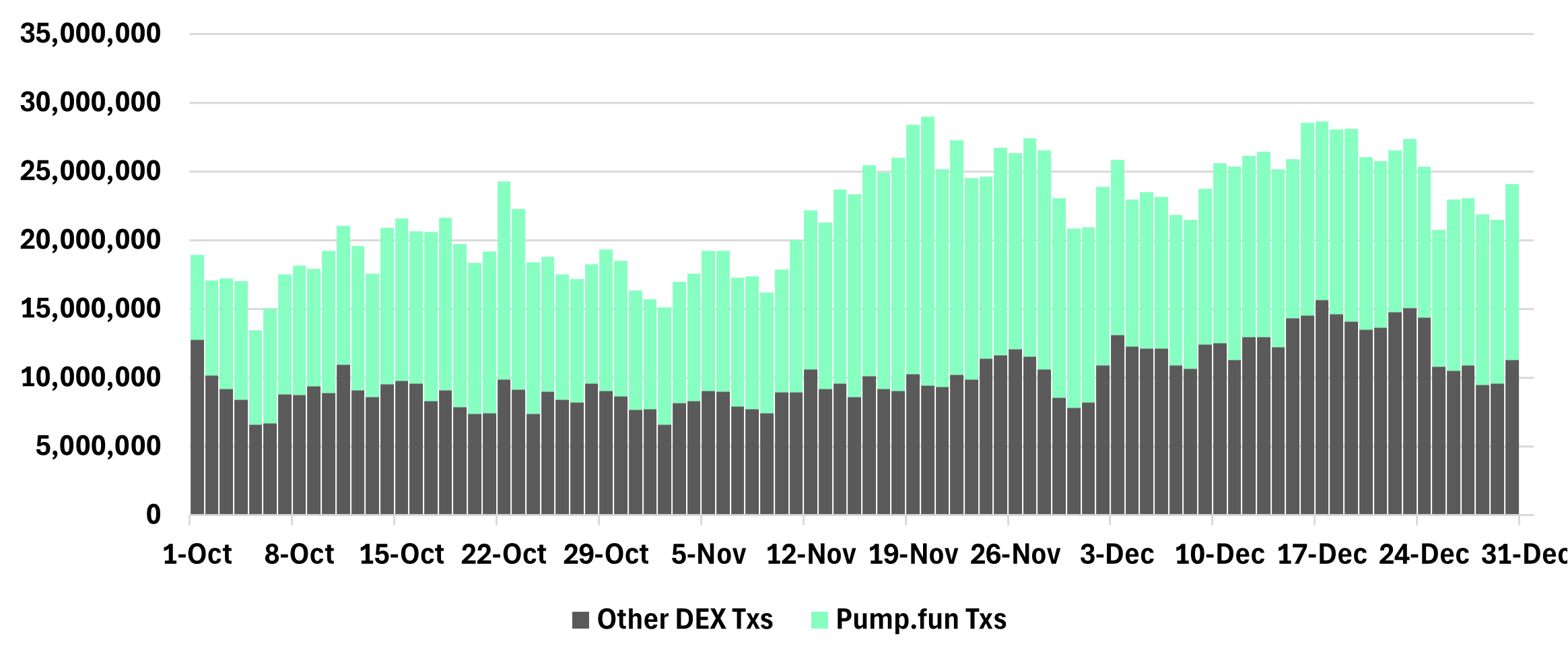}
    \caption{Daily transaction trades of tokens created on Pump.fun.} \label{fig:pftxs}
    
    \vspace{0.5cm}
    
    \includegraphics[width=\linewidth]{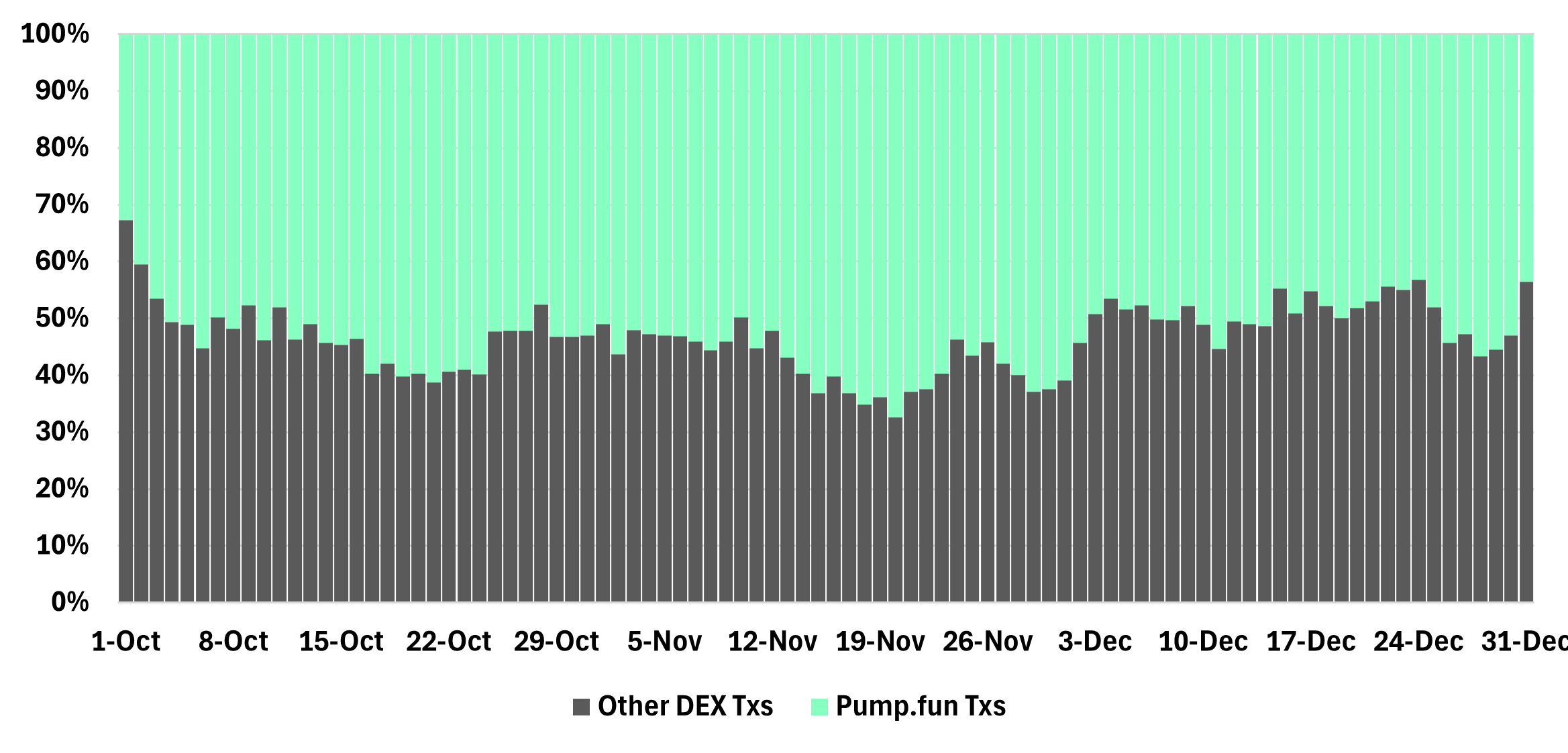}
    \caption{Percentage of daily DEX transactions involving Pump.fun-created tokens.} \label{fig:percPftxs}
\end{figure}

To comprehensively assess Pump.fun's impact on the Solana ecosystem, a metric was employed that encompasses all transactions involving tokens initially created on Pump.fun, irrespective of their current trading platform, as depicted in Figure~\ref{fig:pftxs} and Figure~\ref{fig:percPftxs}.

The methodology for this metric involves:

\begin{enumerate}
    \item Maintaining a comprehensive registry of all tokens created on Pump.fun by analyzing mint instructions in the blockchain data.
    \item Tracking these tokens across all decentralized exchanges (DEXs) by matching token addresses in trade events.
    \item Aggregating transaction data by distinguishing between tokens generated on Pump.fun and those created elsewhere.
\end{enumerate}

This analysis reveals that transactions involving Pump.fun-created tokens constitute between 40\% and 67.4\% of daily DEX transactions. This significant proportion indicates substantial secondary market activity for tokens originating from Pump.fun, underscoring the platform's considerable influence within the Solana ecosystem.

\subsection{Volume Analysis and Retail Focus}
\begin{figure}[!ht]
    \centering
    \includegraphics[width=\linewidth]{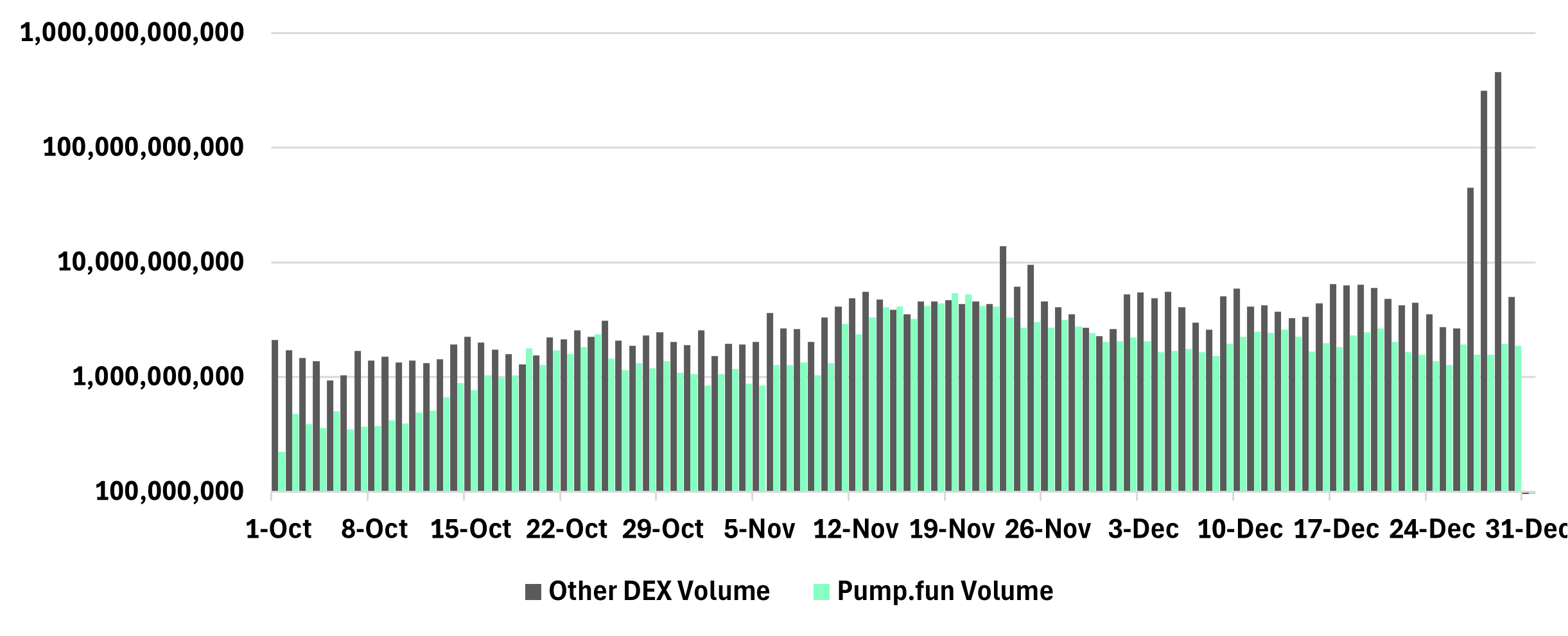} 
    \caption{Daily trading volume of tokens created on Pump.fun (log scale, USD).} \label{fig:pfvol}
    
    \vspace{0.5cm} 
    
    \includegraphics[width=\linewidth]{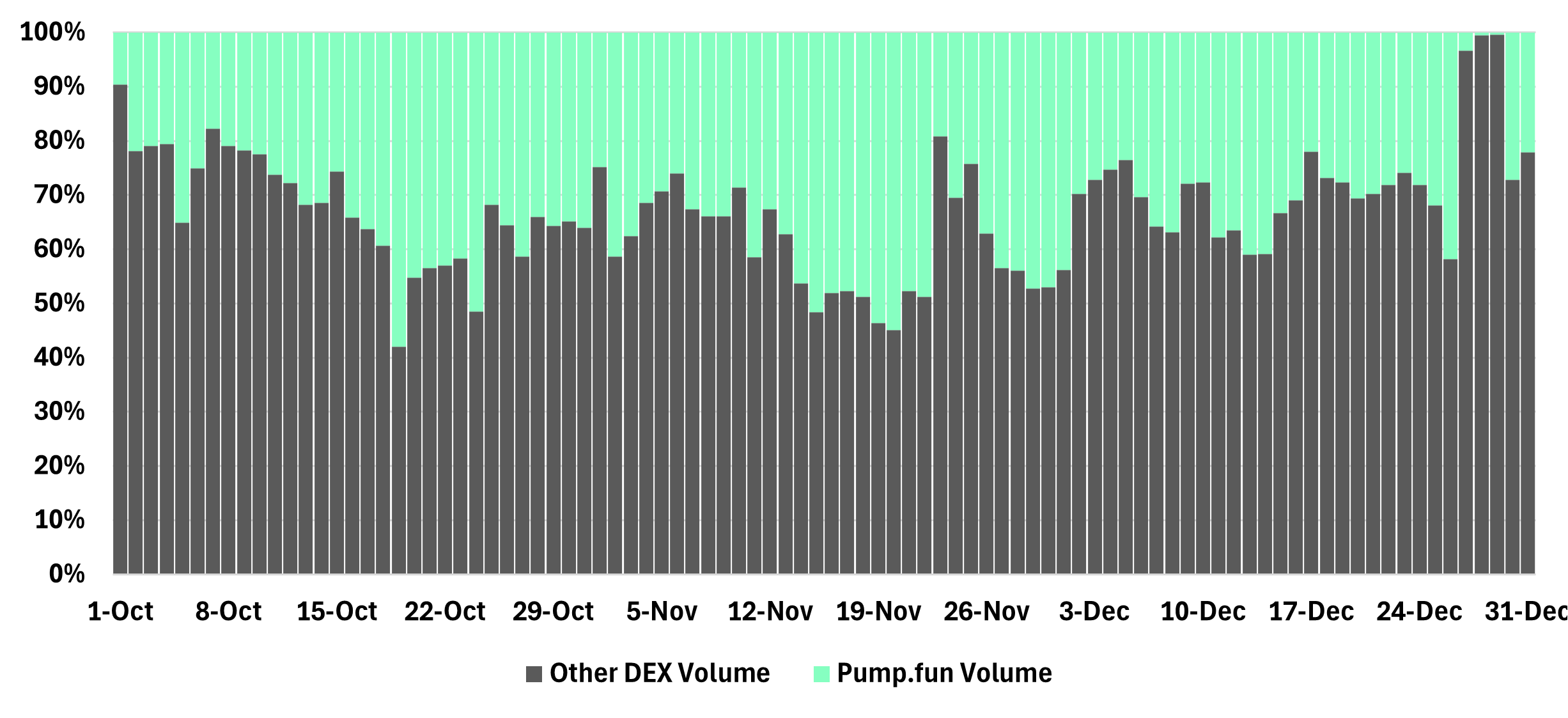} 
    \caption{Percentage of daily trading volume of tokens created on Pump.fun.} \label{fig:percPfvol}
\end{figure}

The volume analysis of the Pump.fun related transactions reveals a dichotomy between transaction counts and trading volumes. As illustrated in Figure~\ref{fig:pfvol} and Figure~\ref{fig:percPfvol}, Pump.fun-related volume demonstrates distinctive characteristics that provide deep insights into the platform's market position and user base composition.

When examining the raw volume data, it is observed that Pump.fun-related trading rarely exceeds 50\% of total DEX volume, despite the platform's dominant position in transaction counts. This apparent discrepancy illuminates a fundamental characteristic of the platform's market position: its pronounced retail orientation. 

\subsection{Token Graduation Analysis}
\begin{figure}[!ht]
    \centering
    \includegraphics[width=\linewidth]{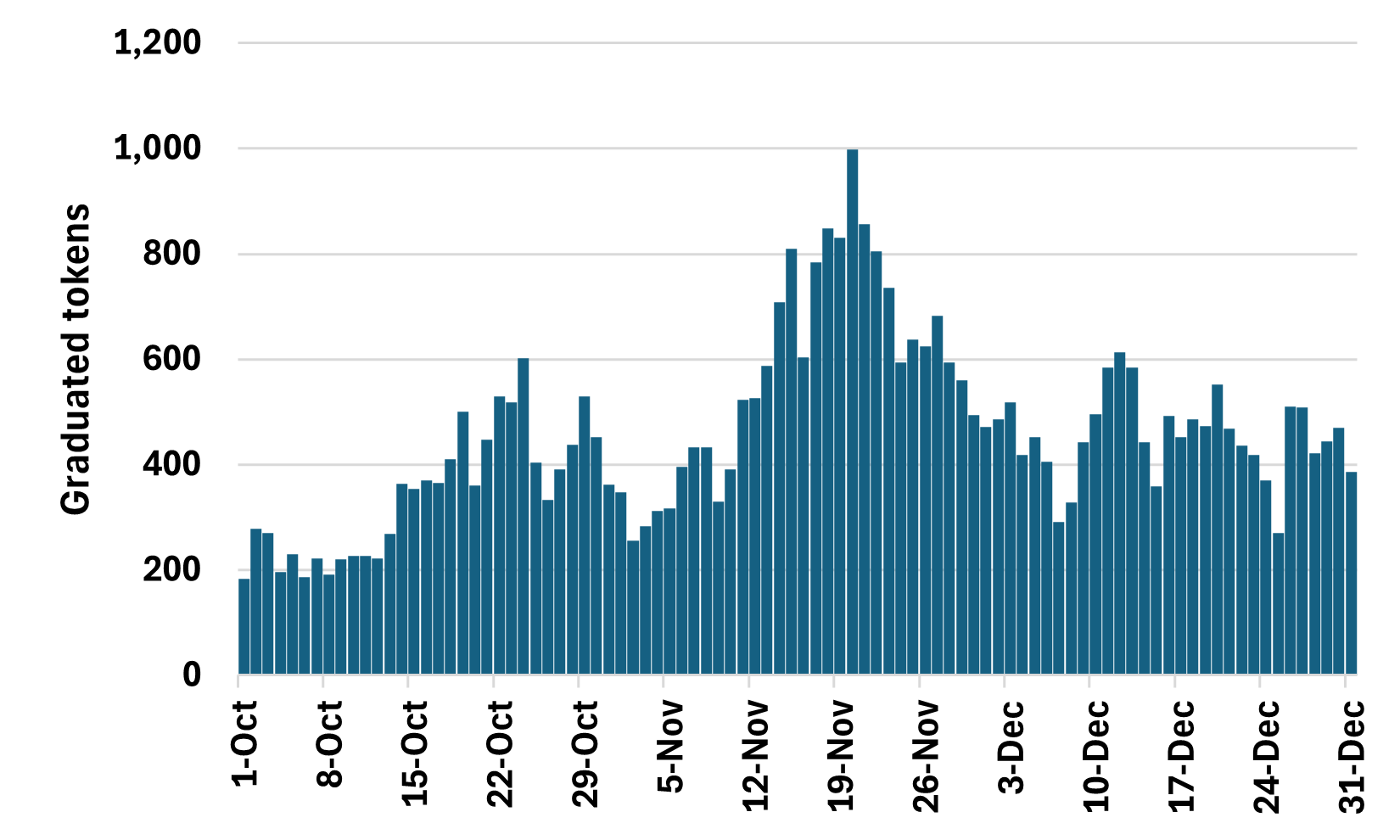} % Sostituisci con il tuo file immagine
    \caption{Daily graduated tokens on Pump.fun.} \label{fig:grtoken}
    
    \vspace{0.5cm} % Spazio tra le immagini
    
    \includegraphics[width=\linewidth]{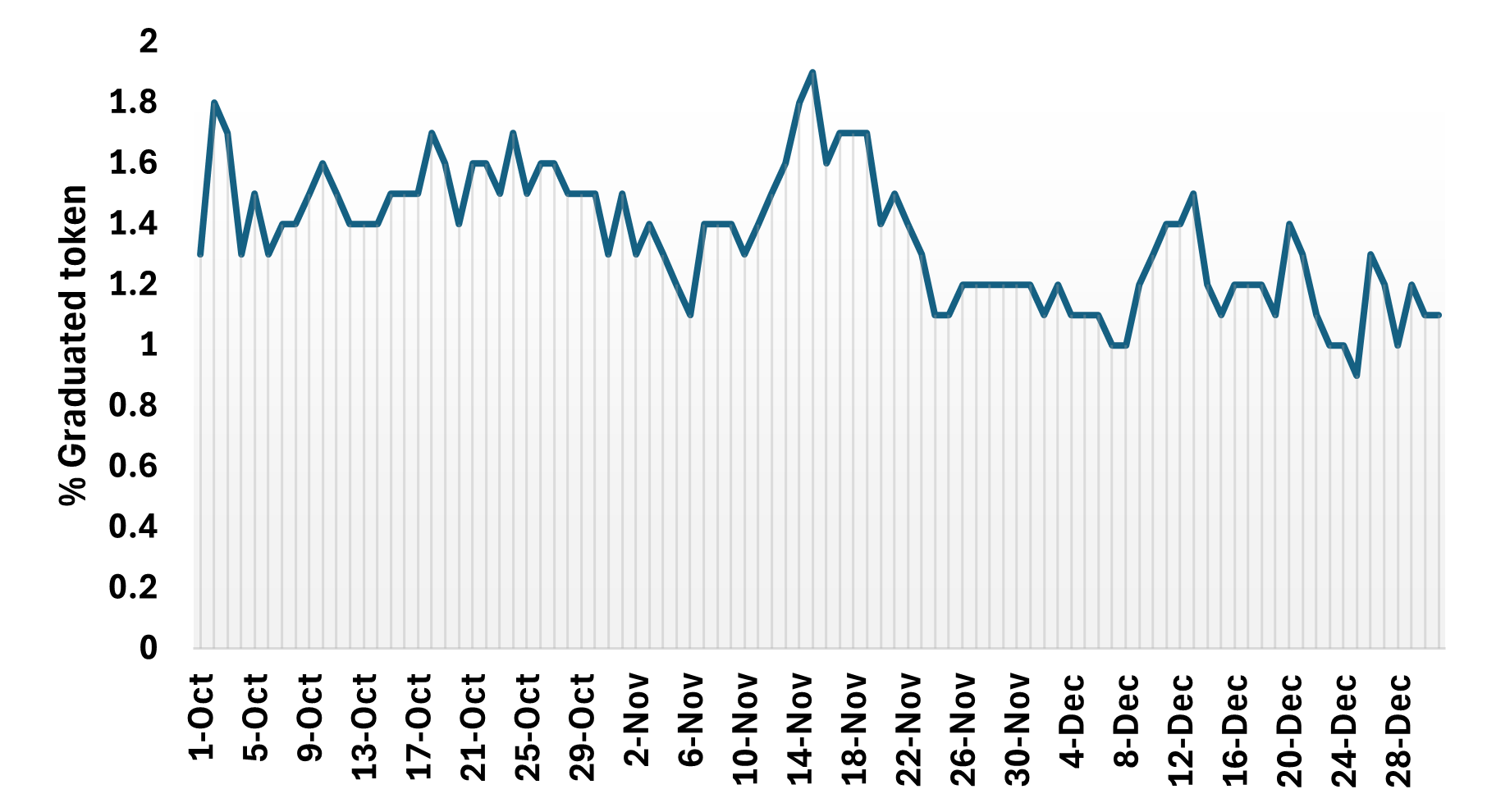} % Sostituisci con il tuo file immagine
    \caption{Percentage of daily graduated tokens relative to daily created tokens on Pump.fun.} \label{fig:percGrtoken}
\end{figure}

The analysis of token graduation rates provides critical insights into the quality and sustainability of tokens created on Pump.fun. 
The SQL query used to analyze token graduation rates identifies tokens created on Pump.fun and tracks their progression to Raydium. It first selects tokens minted on Pump.fun between October 1, 2024, and December 31, 2024. Then, it identifies tokens that have been withdrawn from Pump.fun, indicating their graduation to Raydium. The query aggregates this data to determine the daily number of tokens created on Pump.fun and the daily number of tokens that have graduated to Raydium, calculating the graduation percentage for each day.
By examining graduation events, as visualized in Figures~\ref{fig:grtoken} and~\ref{fig:percGrtoken}, patterns in token lifecycle dynamics and platform efficiency are uncovered.

The graduation process itself merits detailed examination. As discussed in Section \ref{sec:pumpfun}, for a token to graduate to Raydium, it must meet the successful completion of the bonding curve, demonstrating sufficient demand and liquidity.

The analysis of token graduation rates reveals that only a small fraction of tokens created on Pump.fun successfully meet the criteria for listing on Raydium. Specifically, the graduation rate peaks at less than 2\%, indicating that only a limited number of tokens generated on Pump.fun manage to attract sufficient demand and liquidity.

\subsection{User Engagement and Platform Growth}
\begin{figure}[!ht]
    \centering
    \includegraphics[width=\linewidth]{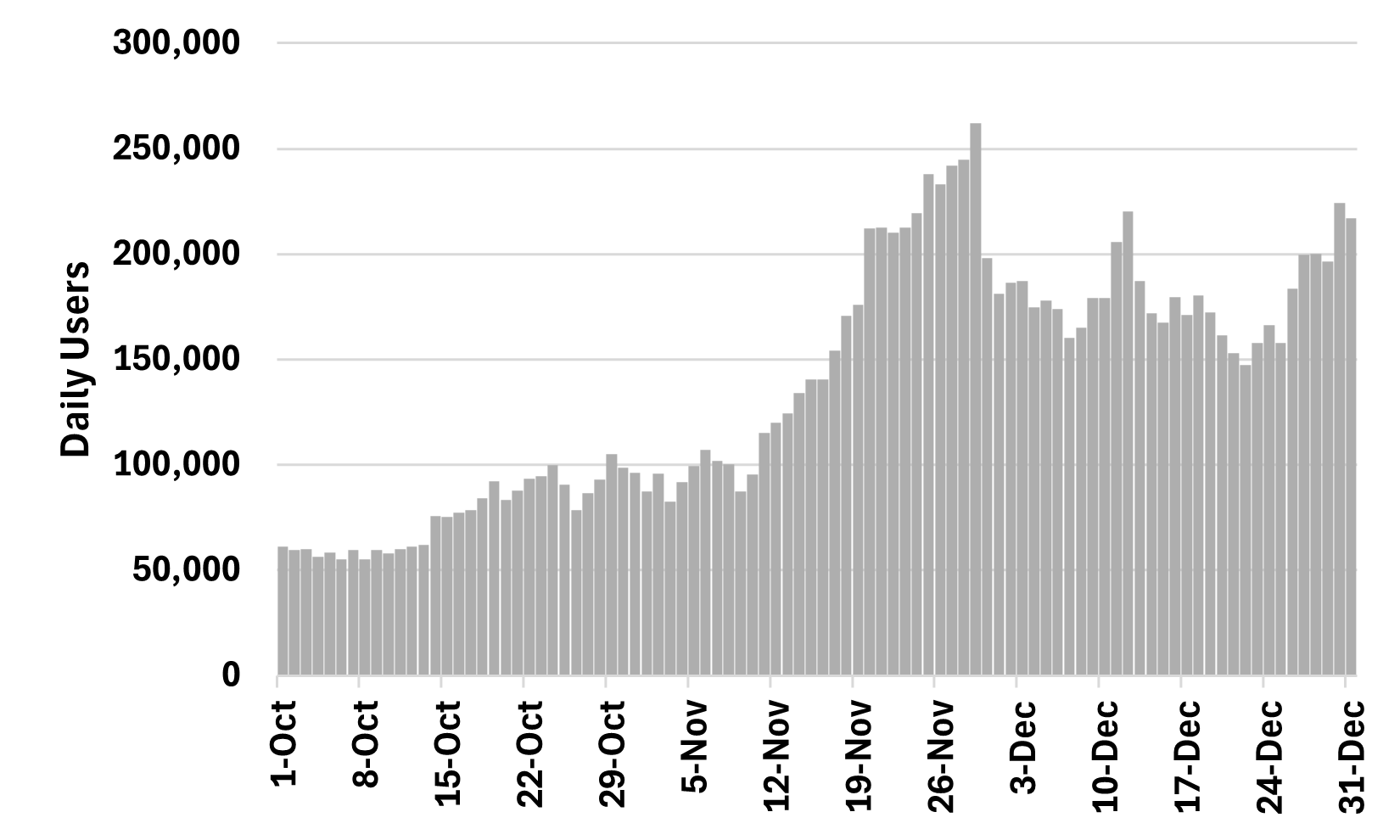} % Sostituisci con il tuo file immagine
    \caption{Daily User growth on Pump.fun.} \label{fig:users}
\end{figure}

The query used for this analysis tracks the daily count of unique user signers for transactions on Pump.fun.
As shown in Figure~\ref{fig:users}, the platform's user base demonstrated extraordinary growth, with unique trading addresses rising from around 60,000 in early October to remarkable peaks of 260,000 in November and 224,000 in December. This figure highlight a significant increase in user activity, suggesting a growing adoption and engagement with the platform. The consistent rise in users over the study period reflects the platform's increasing popularity, making it an increasingly important player in the Solana ecosystem.
  
\section{Economic Implications and Speculative Behavior} \label{sec:speculation}

The analysis of Pump.fun’s ecosystem during Q4 2024 reveals striking patterns of speculative behavior, showcasing the significant influence the platform has had on Solana’s blockchain market. The dominance of Pump.fun in token creation, reaching peaks of 71\% of all Solana tokens minted, signals a major shift in retail trader participation in cryptocurrency markets. This level of engagement suggests that platforms like Pump.fun have democratized the process of token creation, enabling a larger pool of retail investors to engage in speculative activities that were previously dominated by more sophisticated actors.

A key finding in this study is the graduation rate of tokens, with fewer than 2\% reaching sufficient market capitalization for listing on Raydium. This low graduation rate highlights a speculative market structure, where tokens are frequently minted and traded but struggle to achieve long-term sustainability. Despite high transaction frequencies, the relatively low trading volumes suggest that many traders are prioritizing short-term opportunities over long-term value creation. This trend is indicative of a retail-driven market where traders focus on quick profits rather than evaluating the intrinsic value of the tokens they trade. The ecosystem is largely driven by speculative behavior, which thrives on the volatility and uncertainty inherent to the memecoin sector.

Moreover, Pump.fun’s ability to influence 40-67.4\% of all decentralized exchange (DEX) transactions on Solana, even when considering other platforms, underscores how speculation within niche sectors like memecoins can drive broader market dynamics. The platform’s growth, reflected in the substantial increase in daily unique users from 60,000 to 260,000, reveals the mass appeal of speculative opportunities, even when the underlying assets often lack long-term value. These patterns suggest that the platform has capitalized on a growing trend of social coordination among traders, where collective market actions—driven by rumors, trends, and speculation—outweigh traditional market fundamentals.

This retail-driven speculative behavior presents a stark contrast to more traditional forms of blockchain value extraction, such as the MEV (Maximal Extractable Value) strategies analyzed in the Ethereum ecosystem \cite{mancino:offchain_mev:acmdlt:2024}. While MEV relies on technical sophistication, automated systems, and precise transaction ordering to extract value, memecoin speculation through platforms like Pump.fun represents a more accessible, socially-driven form of value extraction. Instead of competing on computational resources and algorithmic efficiency as in MEV scenarios, memecoin traders compete through social coordination, rapid information dissemination, and collective market movements. This fundamental difference highlights how blockchain ecosystems can support diverse forms of speculative activity, each adapted to different participant skillsets and resources.

\section{Related Work} \label{sec:related_work}
This work represents the first comprehensive analysis of a memecoin creation platform and its impact on blockchain market dynamics. While previous research has explored various aspects of memecoins, including their socio-economic implications \cite{krause:memecoin:ssrn:2024} and social media influence \cite{kim:twitter_memecoin:itp:2023}, no study has directly investigated platforms dedicated to memecoin creation. The analysis builds upon research in blockchain market dynamics \cite{mancino:mev_eth:itasec:2023} and DeFi protocols \cite{ozili:defi:jbft:2022}, extending these frameworks to examine how retail-focused platforms are reshaping cryptocurrency market participation through accessible token creation and trading mechanisms.

\section{Conclusion} \label{sec:conclusion}

This research offers a comprehensive examination of the evolving landscape of retail-driven speculative trading on blockchain platforms, with a particular focus on Pump.fun during Q4 2024. The platform's substantial influence in token creation—accounting for up to 71\% of all tokens minted on Solana—highlights the growing trend of memecoin proliferation within the cryptocurrency markets. Retail participants now have greater access to token creation and trading. However, this shift towards accessibility raises important questions about the sustainability of the value being created.

While the platform excels at generating initial trading activity, its low graduation rate—below 2\%—suggests that the ecosystem predominantly supports short-term speculation rather than long-term value creation. This imbalance, coupled with high transaction counts but low trading volumes, indicates that the market structure is driven by frequent, smaller trades, often without sufficient regard for the fundamental value of the tokens involved. This reliance on rapid trading and speculative behavior may present risks to long-term market stability and investor confidence.

The broader implications of this retail-driven speculative behavior are twofold. On the one hand, the growth in unique users and the high number of Pump.fun-related transactions indicate a shift towards more social, crowd-driven forms of market participation. On the other hand, this speculative frenzy risks creating a volatile market that may undermine the broader Solana ecosystem.

Ultimately, the success of Pump.fun in fostering market engagement raises concerns about the long-term impact of platforms focused on short-term speculation. While such platforms may offer new opportunities for retail investors, the risks of market manipulation, volatility, and the creation of low-quality assets must be carefully considered.

The challenges posed by platforms like Pump.fun highlight the broader risks within blockchain-based financial ecosystems. These risks include speculative bubbles, market instability, and the potential for retail investors to be left holding assets that ultimately lack long-term value. As the popularity of such platforms grows, it becomes crucial to develop regulatory and technical frameworks that safeguard against the dangers of unchecked speculation while encouraging responsible market participation.

\bibliographystyle{IEEEtran}
\bibliography{bibliography}

\begin{comment}

\end{comment}

\end{document}